\documentclass[pra,aps,preprint,showpacs,amsmath]{revtex4}
\usepackage{graphicx,bm}

\begin{document}
\title{The effect of dressing on high-order harmonic generation in vibrating H$_2$ molecules}
\author{C.~C.~Chiril\u{a} and M.~Lein}
\affiliation{University of Kassel, Institute of Physics, 
Heinrich-Plett-Stra{\ss}e 40, 34132 Kassel, Germany}

\date{\today}

\begin{abstract}
We develop the strong-field approximation for high-order harmonic generation in hydrogen molecules, including the vibrational motion and the laser-induced coupling of the lowest two Born-Oppenheimer states in the molecular ion that is created by the initial ionization of the molecule. We show that the field dressing becomes important at long laser wavelengths ($\approx 2 \, \mu$m), leading to an overall reduction of harmonic generation and modifying the ratio of harmonic signals from different isotopes.
\end{abstract}

\maketitle

\section{Introduction}

In the high-order harmonic generation (HG) process, an atomic or molecular system irradiated by intense laser light emits high-frequency coherent radiation. The properties of the emitted radiation have led to interesting applications in the past decade. To enumerate just a few, we mention the generation of coherent ultraviolet attosecond pulses \cite{Sansone06, Carrera06, Cao07}, the measurement of vibrational motion in molecules \cite{Baker06}, and tomographic reconstruction of molecular orbitals \cite{Itatani04, Torres07, Patchkovskii07}. This list shows that the main focus of HG has recently extended from atomic systems (mostly rare-gas atoms) to molecules, which possess more degrees of freedom, namely vibration and rotation. They enrich the dynamics and provide additional `control knobs' for HG. By understanding the effects of vibration and rotation on the HG spectra, one is able to manipulate the harmonic radiation.

One of the main theoretical tools in understanding HG is the strong-field approximation (SFA), also known as the Lewenstein model. Originally proposed to study HG in atoms \cite{Lewenstein94}, it was later extended to molecules. It is the quantum-mechanical formulation of the three-step model \cite{Corkum93}, which ascribes HG to a sequence of (i) ionization, (ii) acceleration of the continuum electron, and (iii) recombination. The three-step model has had great success in describing qualitatively the dynamics of harmonic generation. It also predicts correctly the value of the cutoff energy of the emitted harmonic radiation. Regarding the quantitative predictive power of the Lewenstein model for HG in molecules, we note that the model ignores the Coulomb forces acting on the active electron in the continuum. This affects most significantly the region of low harmonics, which is thus not accurately described. For high-order harmonics, the absolute value of the harmonic intensity is usually lower than the value obtained by numerically integrating the time-dependent Schr\"odinger equation (TDSE) (for atoms, see \cite{Chirila04} where such a comparison is made). Nevertheless, in the case of atoms, it was shown that the qualitative behavior of high-order harmonics usually agrees well with the TDSE result \cite{Chirila04}. For molecules, the extra degrees of freedom can hinder the agreement with the exact results. One needs to consider different possible formulations and gauges to decide which one fits better the analysis of a given process. For an extended discussion about the choice of gauge and formulation in the context of molecules, see \cite{Chirila06b,Chirila07}. For example, to describe the two-center interference effects \cite{Lein02}, it is advantageous to use the momentum formulation for the recombination step \cite{Chirila07}. Based on our previous results \cite{Chirila07}, we choose for this work the length-gauge molecular Hamiltonian and the momentum formulation. More systematic, detailed quantitative comparisons to TDSE results are waiting to be performed.

The vibration of the molecular ion, treated in the framework of the Born-Oppenheimer (BO) approximation, was included in \cite{Lein05,Chirila06a}. Only one BO potential surface was taken into account, since at the wavelength of the commonly used Ti:Sapphire laser ($800$ nm), the other potential surfaces are expected to be irrelevant. The reason is that the laser field does not efficiently couple the BO surfaces of the molecular ion during the short time between ionization and recombination. At longer laser wavelengths, available for example at the Advanced Laser Light Source (ALLS) in Canada, the electron excursion times are longer and we expect that the inclusion of excited BO states is important.

 The physical picture we adopt is derived from the simple-man's model applied to molecules, the latter being directly linked to the physical interpretation emerging from the SFA. In this picture, after the active electron has reached the continuum, it does no longer interact with the core, until it returns to recombine. Meanwhile, vibrational motion takes place in the molecular ion. To describe this motion, we consider the two lowest BO potentials in the ion, coupled by the laser field. We investigate two different formulations: (i) the full SFA emerging from the integral equation for the evolution operator and (ii) a simplified model analogous to the atomic simple-man's model, but including the vibration and dressing of the ion. The simple semiclassical model (SM) succeeds in reproducing the general characteristics of the full SFA, while being much less demanding.

In terms of computational effort, the newly proposed SFA model turns out to be very demanding. This is due to solving numerous times the time-dependent Schr{\"o}dinger equation for the vibrational dynamics in the two coupled BO surfaces. The intense numerical effort restricts significantly the parameter space one can explore. To this end, we develop a saddle-point approximation, reducing the computational time dramatically, while accounting for the relevant physical mechanisms. The implementation of this technique is described in detail in the Appendix. 

The paper is organised as follows: in the first part of Section \ref{Theory}, we introduce the new SFA model and briefly discuss the computational details. The second part presents the extended SM model, giving an expression that quantifies the contribution of different electron trajectories (see \cite{Salieres01} and references therein) to the total HG spectrum. Section \ref{Results} discusses the results, for three different laser wavelengths: short ($800$ nm), medium ($1500$ nm), and long ($2000$ nm) wavelengths. The HG spectra for H$_2$ and D$_2$ as well as the ratios of harmonic intensities in the two isotopes are calculated. The characteristics of the results can be well understood from the analysis of the electronic trajectories in the framework of the SM model. The $2000$ nm case fully shows the importance of taking into account more than one BO potential surface. Finally, in the last section we present our conclusions and future perspectives. Atomic units (a.u.) are used throughout this work, unless otherwise specified.

\section{Theory}\label{Theory}

\subsection{The extended strong-field approximation model}
We consider high-order harmonic generation (HG) in the hydrogen molecule, allowing for molecular vibration to take place. The full Hamiltonian includes all Coulomb interactions between the electrons and the protons, but in our model the Coulomb repulsion between electrons is neglected except for using the correct ground-state energy and BO potential of H$_2$. The direction of the molecular axis is kept fixed. In the presence of a linearly polarized electric field ${\mathbf E}(t)$, the length gauge molecule-field interaction is given by $\hat{H}_{\rm{i}}(t) = \hat{H}_{\rm{i1}}(t) + \hat{H}_{\rm{i2}}(t)$, with $\hat{H}_{\rm{i1}}(t)=\mathbf{E}(t) \cdot \mathbf{r}_1$ and $\hat{H}_{\rm{i2}}(t)=\mathbf{E}(t) \cdot \mathbf{r}_2$. Here, $\mathbf{r}_1$ and $\mathbf{r}_2$ are the electron coordinates. The integral equation for the full evolution operator $\hat{U}(t,0)$ reads:
\begin{equation} \label{eq:LS}
\hat{U}(t,0) = \hat{U}_0(t,0) -i \int_0^t dt' \, \hat{U}(t,t') \, [\hat{H}_{\rm{i1}}(t') + \hat{H}_{\rm{i2}}(t')] \, \hat{U}_0(t',0),
\end{equation}
where $\hat{U}_0$ is the evolution operator for the field-free Hamiltonian.

We calculate the total dipole acceleration $\mathbf{a}(t)$ as the time-derivative of the expectation value $\mathbf{P}_{\rm dip}$ of the dipole momentum \cite{Chirila06a} $\hat{\mathbf P}_{\rm dip} = -(\hat{\mathbf P}_1 + 
\hat{\mathbf P}_2)$. Neglecting continuum-continuum contributions, it follows from Eq.~(\ref{eq:LS}):
\begin{equation} \label{eq: dip mom}
\mathbf{P}_{\rm dip}(t) \approx -i \int_0^t dt' 
\langle \Phi^{\rm mol}(t) \vert \hat{\mathbf P}_{\rm dip} \hat{U}(t,t') 
[\hat{H}_{\rm i1}(t') + \hat{H}_{\rm i2}(t')] \vert 
\Phi^{\rm mol}(t')\rangle + \rm{c.c.}
\end{equation}
Here, $\Phi^{\rm mol}(t) = \Phi^{\rm mol} \exp(-i E_0 t)$ and $E_0$ are the molecular ground state (including the vibrational coordinate) and its energy, respectively. Assuming that only one electron can be promoted in the continuum and neglecting the interaction of the continuum electron with the remaining ion, Eq.~(\ref{eq: dip mom}) simplifies to:
\begin{equation} \label{eq: dip mom1}
\mathbf{P}_{\rm dip}(t) \approx 2i \int_0^t dt' 
\langle \Phi^{\rm mol}(t) \vert \hat{\mathbf P}_1 \hat{U}_{\rm 1V}(t,t') \hat{U}_{\rm 2B}(t,t') \hat{H}_{\rm i1}(t') \vert \Phi^{\rm mol}(t')\rangle + \rm{c.c.},
\end{equation}
with the evolution operators $\hat{U}_{\rm 1V}$ and $\hat{U}_{\rm 2B}$ explained below. We used also the fact that the two electrons are equivalent and we neglected the exchange terms \cite{Patchkovskii06,Patchkovskii07,Santra06}. Within the linear combination of atomic orbitals (LCAO) approximation used in this work for describing the electronic orbitals of the initial molecule and of the molecular ion, the exchange term gives rise to an ionization matrix element that is zero for the transition to the electronic ground state of the ion, and non-zero for the transition to the first excited state. The contribution of the latter is negligible compared to the contribution of the direct term (which corresponds to transitions to the electronic ground state of the molecular ion), since the ionization potential for ionization leading into the excited state is bigger than that for the ground state of the molecular ion. 

 Since the exact propagator is unknown, one must resort to approximations. Consequently, in Eq.~(\ref{eq: dip mom1}) the full evolution operator has been factorized in a product of two operators: the Volkov propagator $\hat{U}_{\rm 1V}$ that describes the active electron in the continuum, neglecting the influence of the binding potential and the electron-electron interaction, and the operator $\hat{U}_{\rm 2B}$ that propagates the wave function of the remaining molecular ion. The Volkov propagator can be decomposed spectrally, and the resulting integration over momenta is further simplified by using the saddle-point method \cite{Lewenstein94}. For the propagator $\hat{U}_{\rm 2B}$ we consider only the two lowest-lying BO potential curves. The two potential curves of the molecular ion correspond to the symmetric $\sigma_{\rm g}$ state with electronic wave function $\psi^{\rm ion}_{\rm g}$ and to the anti-symmetric $\sigma_{\rm u}$ state with electronic wave function $\psi^{\rm ion}_{\rm u}$. Including dressing means that the two states are coupled by the external electric field via dipole coupling. At the same time, vibrational wave packets evolve in each BO potential. Finally, the dipole momentum reads:
\begin{eqnarray}\label{eq: dip mom2}
\mathbf{P}_{\rm dip}(t) & \approx & 2i \int_0^t dt' \, \exp(-i S(\mathbf{p}_{\rm s},t,t')) \left[\frac{2\pi}{\epsilon+i(t-t')}\right]^{3/2} \times \\ 
 & {} & \int_0^{\infty}dR \, \chi_0(R) \, \mathcal{L}_{\rm rec}^{T}(R,\mathbf{p}_{\rm s},t) \, \hat{U}_{\rm 2B}(t,t') \, \mathcal{L}_{\rm ion}(R,\mathbf{p}_{\rm s},t') \, \chi_0(R) + c.c., \nonumber
\end{eqnarray}
with $\mathbf{p}_{\rm s}=-\int_{t'}^t dt'' \mathbf{A}(t'') /(t-t')$ being the saddle-point momentum and $\epsilon$ a regularization parameter of the order of unity (we use $\epsilon=1$). The quasiclassical action $S$ is defined as $S(\mathbf{p},t,t') = \int_{t'}^t dt'' \, [\mathbf{p+A}(t'')]^2/2 + \vert E_0 \vert (t-t')$. Here, $\mathbf{A}(t)=-\int_{-\infty}^t dt' \, \mathbf{E}(t')$. The action represents the phase accumulated by the free electron moving under the influence of the external field only (relative to the ground state). The vibrational ground state of the molecule is denoted by $\chi_0(R)$, with $R$ being the internuclear distance.

The ionization and the recombination steps are described by the transition matrix elements $\mathcal{L}_{\rm ion}$ and $\mathcal{L}_{\rm rec}$ (which are vectors of c-numbers):

\begin{equation}\label{eq:ioniz}
\mathcal{L}_{\rm ion}(R,\mathbf{p},t)  =  \left( \begin{array}{c} 
\int\!\!\int \! d\mathbf{r}_1 \, d\mathbf{r}_2  \, \psi_{\rm V}^*(\mathbf{p},\mathbf{r}_1,t) \psi_{\rm g}^{\rm ion}(R,\mathbf{r}_2) \mathbf{E}(t) \cdot \mathbf{r}_1 \psi^{\rm mol}_0(R,\mathbf{r}_1,\mathbf{r}_2) \\
\int\!\!\int \! d\mathbf{r}_1 \, d\mathbf{r}_2  \psi_{\rm V}^*(\mathbf{p},\mathbf{r}_1,t) \psi_{\rm u}^{\rm ion}(R,\mathbf{r}_2) \mathbf{E}(t) \cdot \mathbf{r}_1 \psi^{\rm mol}_0(R,\mathbf{r}_1,\mathbf{r}_2)
\end{array} \right)
\end{equation}

\begin{equation}\label{eq:rec}
\mathcal{L}_{\rm rec}(R,\mathbf{p},t)  =  \left( \begin{array}{c} 
\int\!\!\int \! d\mathbf{r}_1 \, d\mathbf{r}_2 \, \psi_0^{\rm mol}(R,\mathbf{r}_1,\mathbf{r}_2) \hat{\mathbf{P}}_1 \psi_{\rm V}(\mathbf{p},\mathbf{r}_1,t) \psi_{\rm g}^{\rm ion}(R,\mathbf{r}_2) \\
\int\!\!\int \! d\mathbf{r}_1 \, d\mathbf{r}_2 \, \psi_0^{\rm mol}(R,\mathbf{r}_1,\mathbf{r}_2) \hat{\mathbf{P}}_1 \psi_{\rm V}(\mathbf{p},\mathbf{r}_1,t) \psi_{\rm u}^{\rm ion}(R,\mathbf{r}_2)
\end{array} \right).
\end{equation}
In Eqs.~(\ref{eq:ioniz}) and (\ref{eq:rec}), $\psi_{\rm V}(\mathbf{p},\mathbf r,t) = \exp(i(\mathbf{p}+\mathbf{A}(t)) \cdot \mathbf{r})/ (2\pi)^{3/2}$ is the spatial part of a Volkov solution with canonical momentum $\mathbf{p}$ of the electron, and $\psi_0^{\rm mol}$ is the electronic BO ground state of H$_2$. The physical interpretation of the matrix elements in Eqs.~(\ref{eq:ioniz}) and (\ref{eq:rec}) can be given in simple terms. They quantify how much of the initial vibrational wave function $\chi_0$ is transferred on each of the two BO potential surfaces of the molecular ion. The transition is made from the electronic ground state of the molecule to the intermediate state in the HG process via the dipole operator of the active electron. The intermediate state is the state in which the continuum electron is described by a Volkov state and the bound electron is in the $\sigma_{\rm u}$ or $\sigma_{\rm g}$ state of the molecular ion. The recombination matrix elements describe the recombination process of the active electron in the momentum form. We approximate the electronic ground state by $\psi_0^{\rm mol}(R,\mathbf{r}_1,\mathbf{r}_2)=\psi_{\rm g}^{\rm ion}(R,\mathbf{r}_1) \psi_{\rm g}^{\rm ion}(R,\mathbf{r}_2)$ and we use the LCAO approximation for $\psi_{\rm g}^{\rm ion}$ and $\psi_{\rm u}^{\rm ion}$ (only 1s hydrogenic functions are used, see Appendix). In this case, the lower matrix elements in Eqs.~(\ref{eq:ioniz}) and (\ref{eq:rec}) are identically zero, due to symmetry reasons. This would not be the case if another approximation, such as the Heitler-London wave function, was used.

The evolution operator $\hat{U}_{\rm 2B}$ propagates the vibrational wave packets created by ionization in the two potential surfaces of the molecular ion, according to the time-dependent Schr{\"o}dinger equation:
\begin{equation}\label{eq:2L}
i \frac{\partial}{\partial t} \left( \begin{array}{c}
\chi^{\rm ion}_{\rm g}(R) \\
\chi^{\rm ion}_{\rm u}(R)
\end{array} \right) = 
\left( \begin{array}{cc}
-\frac{1}{m_{\rm n}}\frac{\partial^2}{\partial R^2} + V^{\rm ion}_{\rm g}(R) & \mathbf{E}(t)\cdot \mathbf{D}(R) \\
 \mathbf{E}(t)\cdot \mathbf{D}(R) & -\frac{1}{m_{\rm n}}\frac{\partial^2}{\partial R^2} + 
V^{\rm ion}_{\rm u}(R)
\end{array} \right)
\left( \begin{array}{c}
\chi^{\rm ion}_{\rm g}(R) \\
\chi^{\rm ion}_{\rm u}(R)
\end{array} \right),
\end{equation}
where $\mathbf{D}(R)$ is the transition dipole moment between the gerade and ungerade electronic states in the ion, $ V^{\rm ion}_{\rm g,u}(R)$ are the ionic BO energy surfaces, and $m_{\rm n}$ is the mass of one nucleus. The transition dipole moment points along the molecular axis. Its modulus can be well approximated \cite{Bunkin73} by
\begin{equation} \label{eq: dip coup}
\vert \mathbf{D}(R) \vert = 0.4 \, e^{-R} + R/2.
\end{equation}
The numerical propagation for Eq.~(\ref{eq:2L}) is described in detail in \cite{Schwendner97}. At the ionization time $t'$ the initial vibrational wave packets $\chi^{\rm ion}_{\rm g}(R)$ and $\chi^{\rm ion}_{\rm u}(R)$ are given by the initial vibrational ground state $\chi_0$ multiplied by the ionization matrix elements from Eq.~(\ref{eq:ioniz}). The propagation is carried out between the ionization time $t'$ and the recombination time $t$ in Eq.~(\ref{eq: dip mom1}). Hereafter, we refer to the calculation that takes the dipole coupling between the energy surfaces into account as the two-level (2L) calculation. The case when the coupling is neglected is referred to as the one-level calculation (1L), analysed in detail in \cite{Chirila06}.

For large laser wavelengths, the calculation based on Eq.~(\ref{eq: dip mom2}) becomes very time consuming. One solution is to employ the saddle-point method to approximate the integral over the ionization time $t'$. It gives the possibility to study a large parameter space for the laser field, while remaining close to the full SFA results and dramatically reducing of the computational time. The details of the saddle-point method are given in the Appendix. 

\subsection{Simple-man's model including vibration}\label{simple-man}
In order to estimate the importance of field dressing in the process of HG, we investigate the electronic trajectories \cite{Milos02,Sansone04} in the spirit of the simple-man's model \cite{Corkum93}. For each of the classical electronic trajectories, we define a weight that assesses its contribution to the total harmonic spectrum. The trajectories are assumed to start with zero initial velocity at an arbitrary ionization time $t_{\rm i}$. Thereafter the electric field of the laser pulse accelerates the electron and at the moment $t_{\rm r}$ when it returns to the core, the return kinetic energy is calculated. The return kinetic energy plus the ionization potential equals the photon energy of the emitted harmonic radiation.
\begin{figure}[!t]
\centering
\includegraphics[scale=0.4,angle=-90]{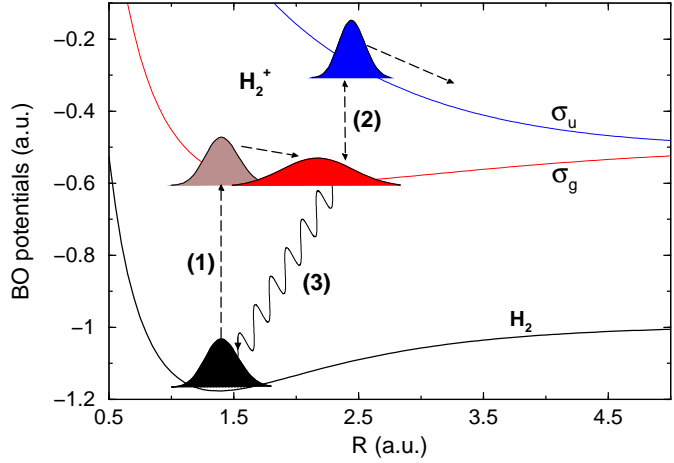} 
\caption{(Color online) Schematic view of the harmonic generation process in H$_2$. Shown are the ground-state BO potential of the H$_2$ and the two lowest BO potentials of H$_2^+$. The three-step process consists of: (1) ionization, (2) evolution of the remaining molecular core as prescribed by the field-dressed ionic $\sigma_{\rm g}$ and $\sigma_{\rm u}$ potential-surfaces while the active electron is driven by the external field, and (3) recombination into the molecular ground state.}
\label{figure1}
\end{figure}
Between $t_{\rm i}$ and $t_{\rm r}$ the vibrational wave packet created in the molecular ion evolves in the two coupled potential surfaces (see Fig.~\ref{figure1}). For the case when only one level is taken into account in the molecular ion, the model was described in \cite{Chirila06}. There, it was shown that the prediction of the model for the ratio of harmonics in D$_2$ and H$_2$ compares very well to the full SFA result at $800$ nm laser wavelength. The simplified model has the advantage that it is easy to implement and it requires very little computational time compared to the full calculation, while containing all the significant physics.

To derive the relevant equations, one relies on applying the saddle point approximation in Eq.~(\ref{eq: dip mom2}), as done for the 1L case in \cite{Chirila06}. The trajectory weight is:
\begin{eqnarray}\label{eq: weight}
w(t_{\rm i},t_{\rm r}) &=& \exp\left(-\frac{2}{3} \frac{(2I_{\rm p})^{3/2}}{\vert E(t_{\rm i})\vert}
\right) \biggl\vert 
\int_0^\infty dR \cos(k_{\rm r} R\cos \theta/2) \times \\
&{}&  \left( \begin{array}{l} \chi_0(R) , 0 \end{array} \right) \hat{U}_{\rm 2B}(t_{\rm r},t_{\rm i})
 \left( \begin{array}{c} \chi_0(R) \\ 0 \end{array} \right) \biggr\vert^2, \nonumber
\end{eqnarray}
where $\theta$ is the orientation angle between the molecular axis and the polarization direction of the electric field $\mathbf{E}(t)$, $E_0$ is the energy of the molecular ground state, and $k_{\rm r}$ is the return velocity of the electron $k_{\rm r}(t_{\rm r},t_{\rm i})=-\int_{t_{\rm i}}^{t_{\rm r}}dt' A(t')/(t_{\rm r}-t_{\rm i})+A(t_{\rm r})$. Equation 
(\ref{eq: weight}) includes essentially all molecular effects on each electron trajectory: the instantaneous ionization rate at time $t_{\rm i}$, the motion of the vibrational wave packets on the coupled BO surfaces in the ion, and the `cos' interference-term. The interference term appears due to the presence of the two molecular sites whose contributions to the harmonic radiation interfere. The wave packet spreading was not included in Eq.~(\ref{eq: weight}), in order to obtain a weight incorporating only the molecular effects. The `vertical' ionization potential $I_{\rm p}$ is defined in the Appendix.

As described in \cite{Milos02,Sansone04}, for electrons that tunnel out during the same optical cycle, there are two types of trajectories: the short trajectories (that last less than about three quarters of the optical period) and the long trajectories. As it will be shown in the following, the influence of the molecular vibration on various trajectories is very different due to the different trajectory durations.

\section{Results}\label{Results}
 As explained in the beginning, we expect that the consequences of vibration and field dressing become more relevant at longer laser wavelengths. To this end, we analyse three different cases. For $800$ nm laser wavelength the effects of field dressing are shown to be negligible. In the intermediate regime, for a laser wavelength of $1500$ nm, the field dressing starts becoming important. Finally, at $2000$ nm, the dressing can no longer be ignored. In our calculations, the laser electric field has a trapezoidal envelope, with $4$ optical cycles turn-on and turn-off, and $6$ optical cycles of constant amplitude. The laser field is linearly polarized, and unless specified, the molecule is oriented parallel to the laser polarization direction ({\it i.e.}, we consider aligned molecules). The laser intensity is $5\times 10^{14}$ W/cm$^2$, unless stated otherwise. All results in this section have been obtained using the saddle-point approximation (see Appendix), since for long wavelengths, the computational times for the fully numerical SFA are prohibitive. Except in the range of low harmonics, the saddle-point 1L calculations were found in excellent agreement with the harmonic spectra calculated by full numerical integration with the field dressing neglected; for more details, see Appendix. The propagation of the vibrational wave packets has been carried out on a grid. The grid method is slightly more accurate than the eigenfunction decomposition method employed in \cite{Chirila06} for the 1L SFA. 

\subsection{$800$ nm laser field}
\begin{figure}[!t]
\centering
\includegraphics[scale=0.45,angle=-90]{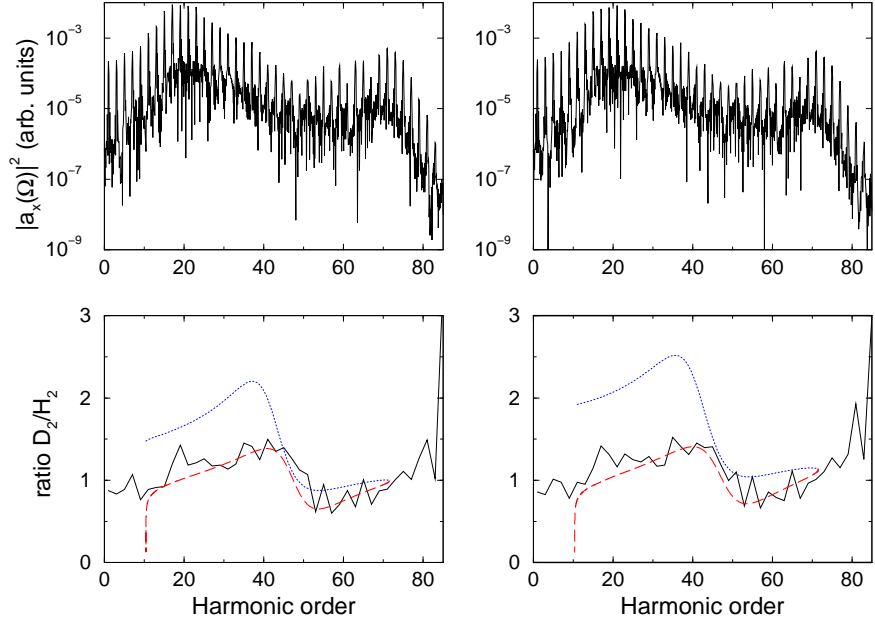} 
\caption{(Color online) Harmonic generation using $800$ nm laser pulses. Left column: 1L case, right column: 2L case. The upper row shows the harmonic intensities for the H$_2$ molecule. The lower row shows the harmonic ratio D$_2$/H$_2$ (black, continuous curve), and the ratio predicted if one takes into account the short trajectory only (red, dashed curve) or the long trajectory only (blue, dotted curve).}
\label{figure2}
\end{figure}

Figure \ref{figure2} compares the 1L and the 2L calculations. The reader should ignore the region of harmonic orders below $\approx 15$. In the saddle-point approximation, the electronic trajectories with small travel times are not accounted for, so that the low-order harmonics are not treated accurately (see comment at the end of the Appendix).

As expected, the difference caused by the field dressing is very small.
This can be understood, since the typical travel time in the harmonic generation process is of the order of one optical cycle, during which the transfer of the vibrational wave packet on the excited potential surface of the molecular ion is negligible. A clear difference between D$_2$ and H$_2$ appears when the ratio of harmonic intensities is taken (see bottom row). The ratio exhibits a strong variation around harmonic order $50$. Comparing to the harmonic spectrum, this is the region where an interference minimum \cite{Lein02} in the harmonic emission occurs. Due to the different masses of the isotopes causing different speeds of nuclear motion, the position of the interference minimum slightly changes from one isotope to the other. The cutoff of the electronic trajectories approximates well the quantum-mechanical cutoff, as expected. For the trajectories, we show only the pair of short and long trajectories that has the biggest contribution to the harmonic spectrum. It is evident that the short trajectory is the one that reproduces well the exact ratio (see Fig.~\ref{figure2}, bottom row) \cite{Chirila06}. We note that the part of the harmonic spectrum with frequencies below the value of the `vertical' ionization potential (see Appendix) is not accessible for the classical trajectory analysis.

\begin{figure}[!t]
\centering
\includegraphics[scale=0.45,angle=-90]{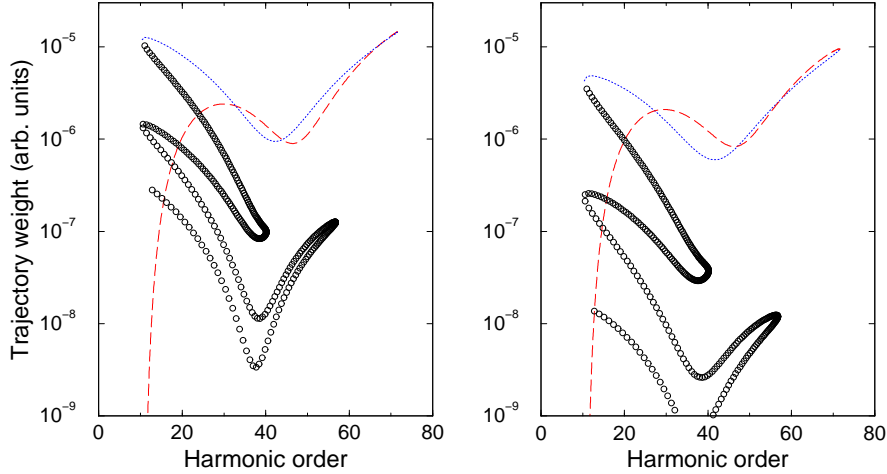} 
\caption{(Color online) Trajectory weights for $800$ nm laser pulses (H$_2$ only). Left panel: the trajectory weights for the 1L calculation. Right panel: the trajectory weights for the 2L calculation. The red, dashed curves are used for the short trajectory and the blue, dotted curves are for the long trajectory. The black circles correspond to the longer trajectories. The trajectories shown by the dashed and the dotted lines are used to calculate the ratios depicted by the same curves in Fig.~\ref{figure2}.}
\label{figure3}
\end{figure}

To deeper understand the role of different electron trajectories in the harmonic spectrum, Fig.~\ref{figure3} shows their weights [see Eq.~(\ref{eq: weight})]. 
As it can be seen, different pairs of trajectories contribute to different regions of the harmonic spectrum. Each pair consists of a short and a long trajectory. The pair with the shortest excursion time was used to calculate the ratios shown in Fig.~\ref{figure3}, since this pair has the biggest contribution. This explains its success in reproducing satisfactorily the exact ratio. The shorter trajectories are not strongly affected by the laser coupling at this wavelength, while the pairs with longer travelling time can be strongly affected by the field coupling (compare the lowest pairs in the left and right panel of Fig.~\ref{figure3}). These pairs do not contribute significantly to the total spectrum.
\subsection{$1500$ nm laser field}
\begin{figure}[!t]
\centering
\includegraphics[scale=0.45,angle=-90]{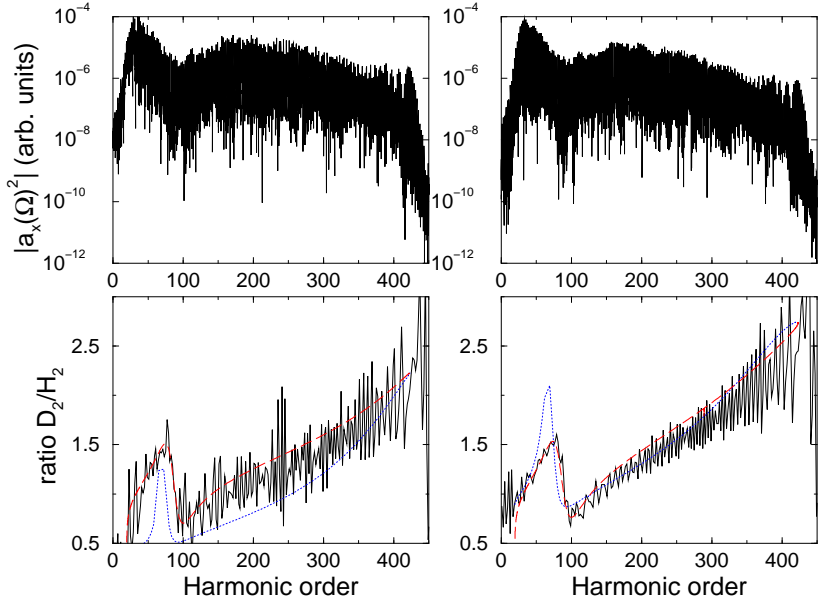} 
\caption{(Color online) Harmonic generation using $1500$ nm laser pulses. Left column: 1L case, right column: 2L case. The upper row shows the harmonic intensities for the H$_2$ molecule. The lower row shows the harmonic ratio D$_2$/H$_2$ (black, continuous curve), and the ratio predicted if one takes into account the short trajectory only (red, dashed curve) or the long trajectory only (blue, dotted curve).}
\label{figure5}
\end{figure}

\begin{figure}[t]
\centering
\includegraphics[scale=0.4,angle=-90]{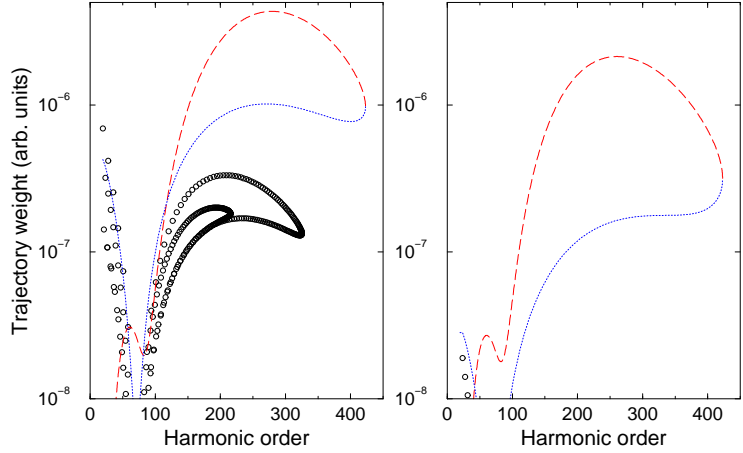} 
\caption{(Color online) Trajectory weights for $1500$ nm wavelength (H$_2$ only). Left panel: the trajectory weights for the 1L calculation. Right panel: the trajectory weights for the 2L calculation. The red, dashed curves are used for the short trajectory and the blue, dotted curves are used for the long trajectory. The black circles correspond to the longer trajectories. The trajectories shown by the dashed and the dotted lines are used to calculate the ratios depicted by the same curves in Fig.~\ref{figure5}.}
\label{figure6}
\end{figure}

 Due to the longer duration of an optical cycle for $1500$ nm wavelength, one expects to see stronger signatures of the field dressing in the harmonic spectra. This is indeed evident in the results shown in Fig.~\ref{figure5}.
Comparing the ratio of the harmonic signal (bottom row of Fig.~\ref{figure5}) for the 1L calculation and the 2L calculation, the short trajectory is again able to reproduce quantitatively the full results. In the 2L case, the ratio exhibits a somewhat smoother variation with the harmonic order. We understand this by analysing the behavior of the trajectory weights in Fig.~\ref{figure6}.
Here, there is a significant influence of the field dressing on the trajectories. Namely, when the field coupling is included, the short trajectory is the one least affected, while the long one and the remaining less-contributing pairs become strongly damped. In such a case, the quantum interference effects between trajectories \cite{Milos02,Sansone04} have a smaller impact on the HG spectrum, so that the ratio becomes smoother. Again, note the strong variation of the ratio below harmonic order $150$, which is related to the position of the interference minimum \cite{Lein02}, as discussed above.

\subsection{$2000$ nm laser field}
Finally, we study the case when the laser wavelength is large enough to allow the field dressing effects to fully manifest themselves. We show results for two laser intensities, in order to assess the influence of the field strength on the coupling of the BO surfaces. 
 
\begin{figure}[!t]
\centering
\includegraphics[scale=0.55,angle=-90]{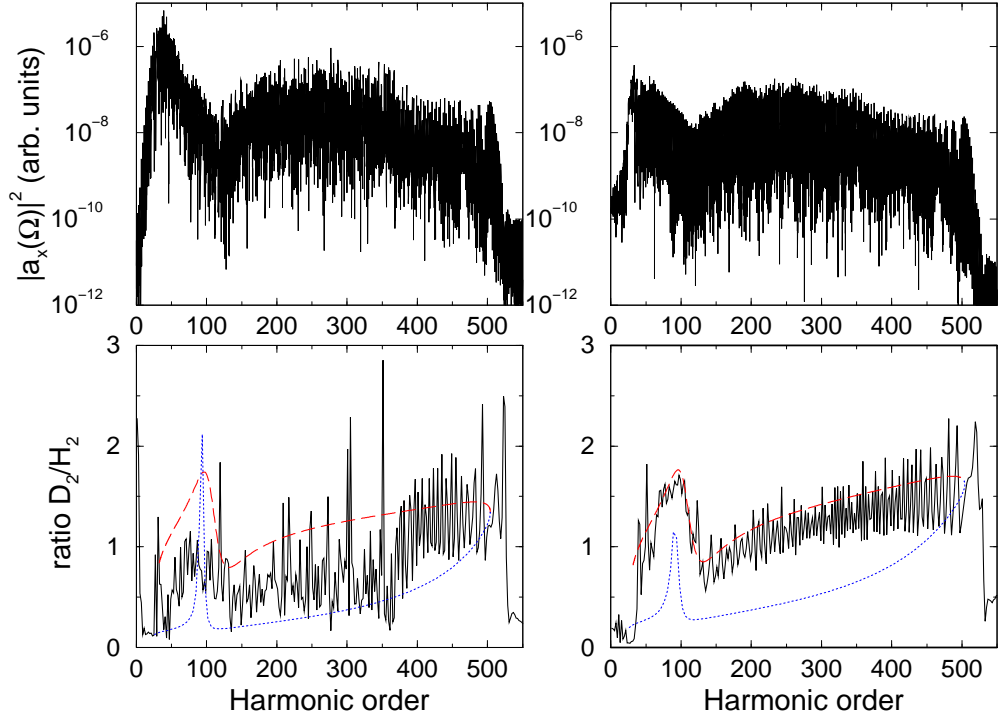} 
\caption{(Color online) Harmonic generation using $2000$ nm laser pulse with $2.5 \times 10^{14}$ W/cm$^2$ intensity. Left column: 1L case, right column: 2L case. The upper row shows the harmonic intensities for the H$_2$ molecule. The lower row shows the harmonic ratio D$_2$/H$_2$ (black, continuous curve), and the ratio predicted if one takes into account the short trajectory only (red, dashed curve) or the long trajectory only (blue, dotted curve).}
\label{figure7}
\end{figure}

For the laser intensity I=$2.5 \times 10^{14}$ W/cm$^2$, Fig.~\ref{figure7} compares the HG spectra for the 1L and the 2L case (top row).
The reduction in the harmonic intensity is clearly visible on the scale of the graphs. Thus, one concludes that the field coupling can not be neglected at long wavelengths. 

\begin{figure}[!t]
\centering
\includegraphics[scale=0.4,angle=-90]{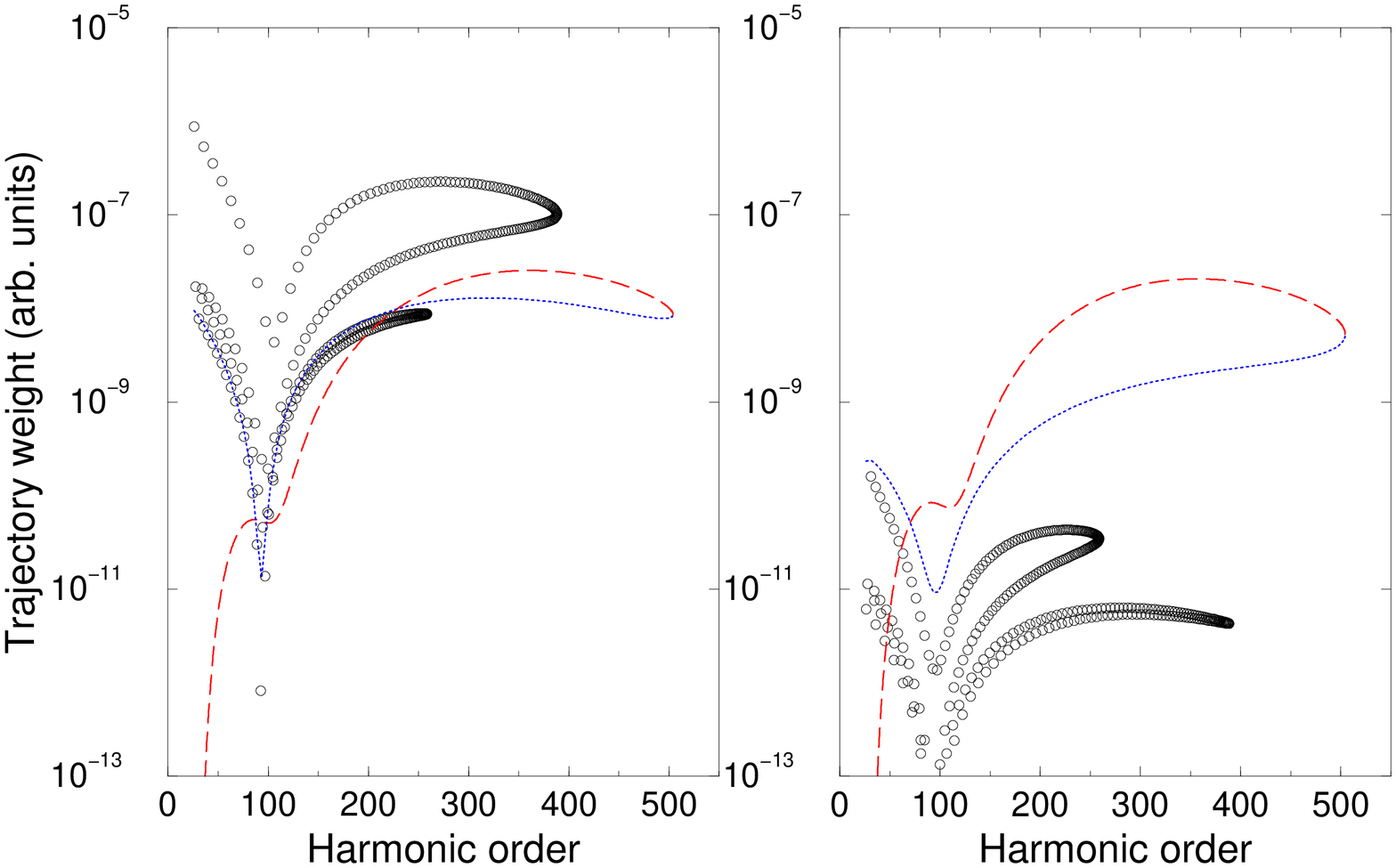} 
\caption{(Color online) Trajectory weights for $2000$ nm wavelength and $2.5 \times 10^{14}$ W/cm$^2$ intensity (H$_2$ only). Left panel: the trajectory weights for the 1L calculation. Right panel: the trajectory weights for the 2L calculation. The red, dashed curves are used for the short trajectory and the blue, dotted curves are for the long trajectory. The black circles correspond to the longer trajectories. The trajectories shown by the dashed and the dotted lines are used to calculate the ratios depicted by the same curves in Fig.~\ref{figure7}.}
\label{figure8}
\end{figure}

The HG spectrum for the 2L case shows less trajectories interference and hence a smoother shape of the spectrum envelope . This feature is explained by the simple-man's analysis (Fig.~\ref{figure8}), which shows that the contribution of the longer trajectories to the HG spectrum is strongly attenuated due to the larger time the electrons spends in the continuum. For the same reason, the SFA harmonic ratio for the 2L case (bottom right panel of Fig.~\ref{figure7} ) agrees with the SM ratio much better than for the 1L case (bottom left panel of Fig.~\ref{figure7}).
In the 1L case (left panel of Fig.~\ref{figure8}), the long trajectories with duration more than one cycle contribute significantly to the spectrum, at least for photon energies below the $\approx 400^{\rm th}$ harmonic order. At higher harmonic orders, there is only one trajectory pair that contributes to the HG spectrum. Consequently, the SM model succeeds in reproducing well the full SFA ratio (bottom left panel of Fig.~\ref{figure7}).
 
Both the 1L and the 2L HG spectra show a minimum around the $120^{\rm th}$ harmonic order. As discussed above, this minimum is caused by the interference of the emitted harmonic radiation from the two molecular sites in D$_2$ \cite{Lein02}. Noticeably, the minimum appears clearly in both the full SFA and the SM harmonic ratios. The presence of the cosine interference term in the expression for the trajectory weight given by Eq.~(\ref{eq: weight}) is essential for reproducing this minimum.

\begin{figure}[!t]
\centering
\includegraphics[scale=0.45,angle=-90]{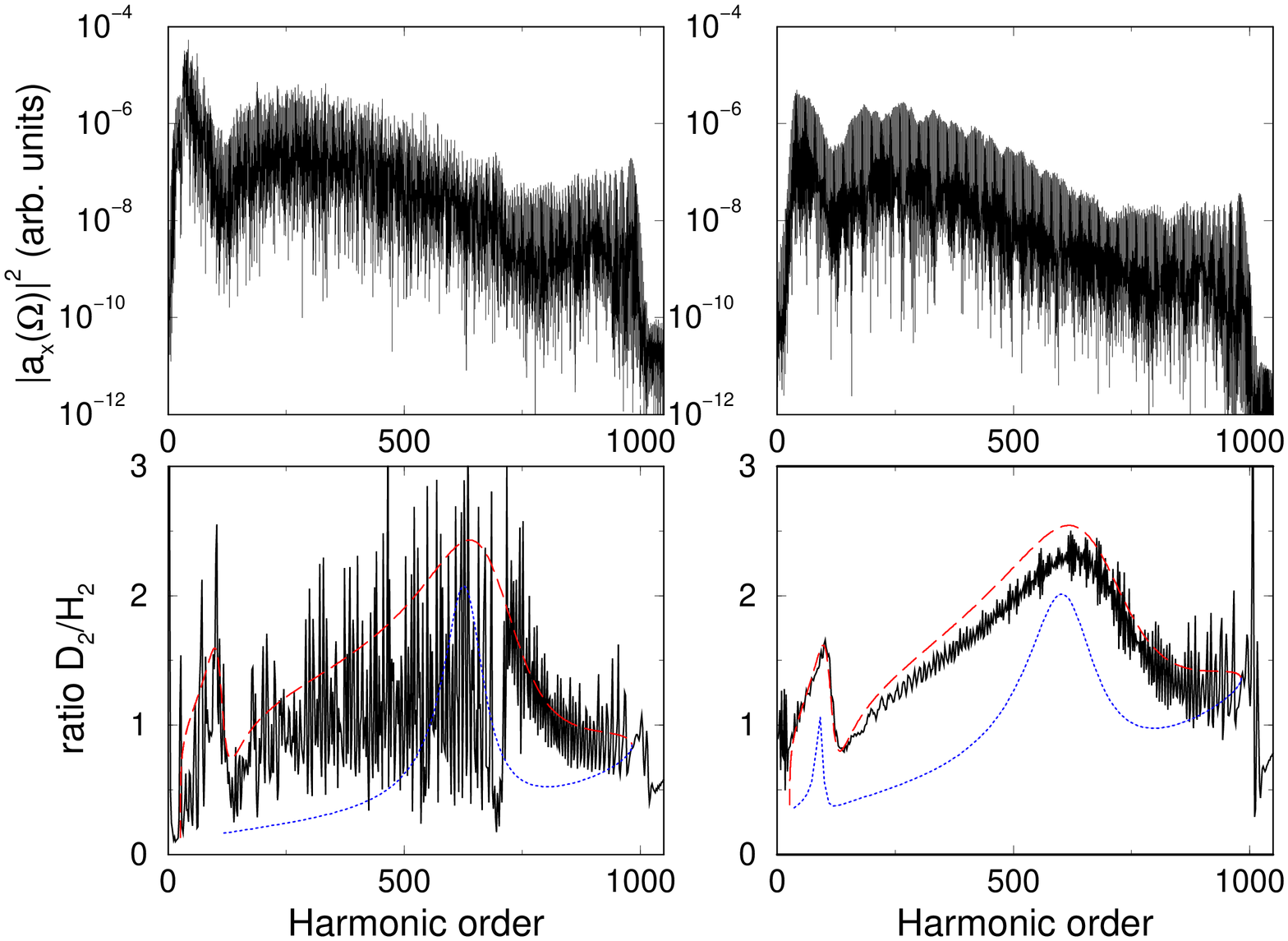} 
\caption{(Color online) Harmonic generation using $2000$ nm laser pulses with $5 \times 10^{14}$ W/cm$^2$ intensity. Left column: 1L case, right column: 2L case. The upper row shows the harmonic intensities for the H$_2$ molecule. The lower row shows the harmonic ratio D$_2$/H$_2$ (black, continuous curve), and the ratio predicted if one takes into account the short trajectory only (red, dashed curve) or the long trajectory only (blue, dotted curve).}
\label{figure9}
\end{figure}

The effects of field coupling become even more apparent at higher laser intensity. In the following, we employ the value I=$5 \times 10^{14}$ W/cm$^2$, used also for the shorter laser wavelengths. From Fig.~\ref{figure9}, by comparing the HG spectra for the 1L and the 2L cases, the difference between the 1L and the 2L results is noticeable.

\begin{figure}[!t]
\centering
\includegraphics[scale=0.4,angle=-90]{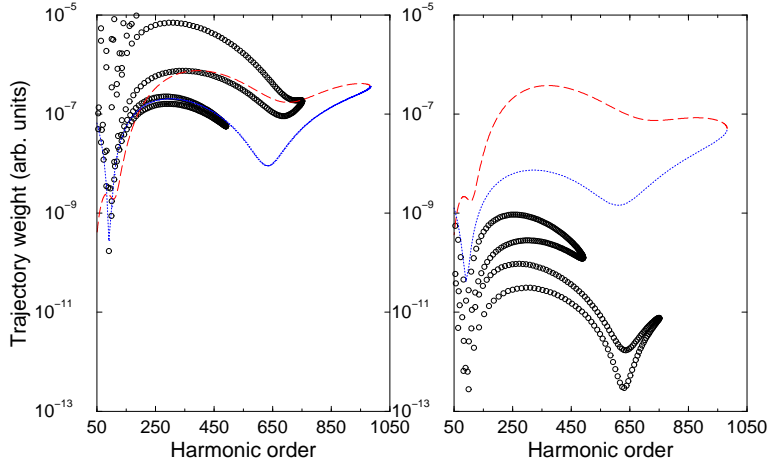} 
\caption{(Color online) Trajectory weights for $2000$ nm wavelength and $5 \times 10^{14}$ W/cm$^2$ intensity. Left panel: the trajectory weights for the 1L calculation. Right panel: the trajectory weights for the 2L calculation.  The red, dashed curves are used for the short trajectory and the blue, dotted curves are for the long trajectory. The black circles correspond to the longer trajectories. The trajectories shown by the dashed and the dotted lines are used to calculate the ratios depicted by the same curves in Fig.~\ref{figure9}.}
\label{figure10}
\end{figure}

In the 1L case, the ratio is not well reproduced by the short trajectory, except beyond the $700^{\rm th}$ harmonic. The reason is analogous to the lower-intensity case, as becomes clear when one analyses the trajectories weights shown in the left panel of  Fig.~\ref{figure10}. As more trajectories contribute to harmonic orders below $700$, quantum interference takes place. Consequently, one single trajectory is not enough to describe accurately the HG spectrum. In contrast, for harmonics of order higher than $700$, there is only one short trajectory that dominates the HG spectrum. This clearly explains the agreement between the full result and the predictions of the SM model in this region. We have checked that these findings remains valid when additional factors due to wave-packet spreading are taken into account.

The situation for the 2L case is strikingly different from the 1L case. Namely, the ratio is much smoother, with reduced signs of interference. Such behavior is caused by the fact that most of the trajectories that would normally contribute to the spectrum become less important. The classical analysis shows that this is indeed the case (see Fig.~\ref{figure10}). The SM analysis shows that even the short trajectories are affected strongly by the field coupling. As a general characteristic of large wavelengths, the harmonic spectra tend the be `smoothed out' by the field dressing ({\it i.e.}, less interferences). In the 1L case, for an extended part of the spectrum, there are many 
trajectories that contribute with comparable weights, so that the trajectory interference in the harmonic signal is strong. This makes the SM model inapplicable to explain the ratio.
When the field dressing is included, the long trajectories are damped, hence the interference is almost absent. This reveals an important feature of the harmonic ratio, namely a maximum around harmonic order $600$, which is due to second order destructive two-center interference in H$_2$.

\begin{figure}[!t]
\centering
\includegraphics[scale=0.5,angle=-90]{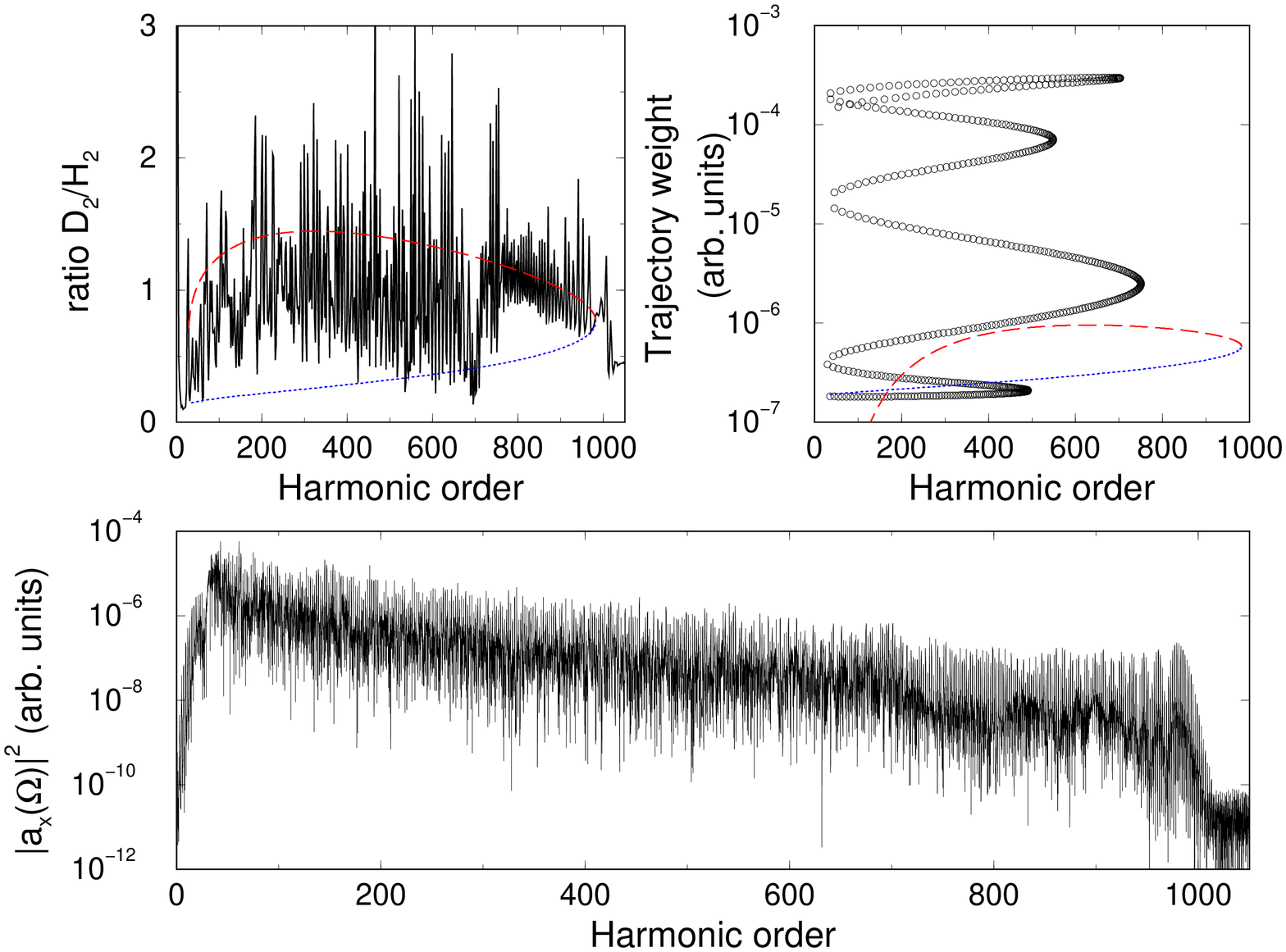} 
\caption{(Color online) Harmonic generation using $2000$ nm laser with $5 \times 10^{14}$ W/cm$^2$ interacting with molecules perpendicular to the laser polarization direction ($\theta=90^{\circ}$). Top row: Harmonic ratio D$_2$/H$_2$ compared to the SM ratios for shortest trajectories (left) and the trajectory weights for H$_2$ from the simple-man's model (right). The red, dashed curves are used for the short trajectory and the blue, dotted curves are for the long trajectory. The black circles in the top-right panel correspond to the longer trajectories. Bottom: harmonic spectrum for H$_2$.}
\label{figure11}
\end{figure}

In the remaining part of this section we consider the case when the molecule is perpendicular to the laser polarization direction ($\theta=90^{\circ}$). The laser intensity is the $5 \times 10^{14}$ W/cm$^2$ as in Fig.~\ref{figure9}. For this orientation, both dressing and two-center interference \cite{Lein02} are absent. This is confirmed by the flat HG spectrum (lower part of Fig.~\ref{figure11}).
The top row of Fig.~\ref{figure11} shows the harmonic ratio obtained from the SFA compared to the SM ratio (left panel), and the trajectory weight (right panel). The SM model approximates the harmonic ratio well for harmonic orders higher than $700$ for the same reason as in the 1L case for parallel alignment (see trajectory 
weights in the upper-right panel of Fig.~\ref{figure11}). 

An important observation is that the shortest electron trajectory pair does not have the highest weight for the lower harmonics (upper-right panel of Fig.~\ref{figure11}). We have found numerically that the weight of the shortest pair increases with decreasing wavelength and becomes dominant at wavelengths lower than $\approx 1300$ nm. This behavior is related directly to the temporal shape of the vibrational autocorrelation function \cite{Chirila06a}. It has a maximum at the time of the vibrational period of the ion, giving increased weight to long trajectories with suitable duration. Our results show that this effect plays a role only when the dressing is absent.

\section{Conclusions}
In this work, we have analysed the possibility to include field dressing in the strong-field approximation for harmonic generation in H$_2$ molecules. Previously, the vibration of the molecular ion formed upon ionization was considered to take place on the lowest BO potential surface only \cite{Chirila06}. Here, we take into account two potential surfaces, coupled by the external field via the dipole interaction. Such a modification proves to be essential at long laser wavelengths.

To study the effect of field dressing, we use an extension of the Lewenstein model which includes fully quantum-mechanical vibrational motion of the molecular ion in the field-coupled lowest two BO surfaces. The resulting expression for the electronic dipole momentum turns out to become numerically demanding even at the moderate wavelength of $800$ nm. The reason is that a large number of one-dimensional time-dependent Schr{\"o}dinger equations have to be solved. At longer wavelengths, the calculation becomes seriously prohibitive. Therefore, we applied the saddle-point method successfully, replacing the time integration by a sum over a few relevant terms.

We have investigated three different laser wavelengths, $800$, $1500$, and $2000$ nm. For the $800$ nm laser field, the effects of the laser dressing can be safely neglected. At $1500$ nm wavelength and more strongly at $2000$ wavelength, the effects of the field dressing manifest themselves in the harmonic spectrum, and more prominently in the ratio of harmonic intensities for D$_2$ and H$_2$. The field coupling has the effect of `smoothing' the interferences in the harmonic spectrum, by lowering dramatically the contribution of the long trajectories to the HG spectrum. Consequently, only the shortest trajectory contributes significantly to the spectrum. At $2000$ nm, we found that this effect leads to a much clearer observation of two-center interference in the ratio D$_2$ vs.~H$_2$.

The proposed model could prove its usefulness in interpreting the experimental data available from the use of the recently available long laser-wavelength sources and for uncovering new properties of the harmonic radiation in molecular systems.

This work was supported by the Deutsche Forschungsgemeinschaft.

\appendix*
\section{Saddle-point approximation}\label{appendix}
The saddle-point method for the study of harmonic generation has been successfully applied in the context of the strong-field approximation ({\it e.g.}, for atoms see \cite{Milos02}, and for non-vibrating molecules see \cite{Faria07}), due to the strongly oscillating phase factor $\exp(-iS)$ in the expression of the dipole moment [see Eq.~(\ref{eq: dip mom2})]. In the present work, we apply the saddle-point method to approximate the 
integral over the ionization time $t'$ in  Eq.~(\ref{eq: dip mom2}). Compared to atoms and non-vibrating molecules, one encounters the additional problem that the part of the integrand which describes the molecular vibration can also oscillate with the integration variable. This oscillation cannot be calculated analytically. If the oscillatory vibrational part was known analytically, one could calculate the saddle points exactly. Based on our previous analysis of the 1L case \cite{Chirila06}, we concluded that it is possible to isolate the desired oscillatory part. In order to do this, we make use of the fact that the main contribution to the electronic dipole moment comes from transitions between the electronic ground state of the initial molecule and that of the molecular ion. Consequently, we subtract from the energy curves of the molecular ion the quantity $\delta V = V_{\rm g}^{\rm ion}(\bar R)$, with $\bar R$ the average internuclear distance in the vibrational ground state of the neutral molecule. This means that, the propagation in Eq.~(\ref{eq:2L}) is done with the energy curves re-defined as $V_{\rm g,u}^{\rm ion} = V_{\rm g,u}^{\rm ion}-\delta V$, while a term $(t-t')\delta V$ has to be added to the action in Eq.~(\ref{eq: dip mom2}). Combined with the ground-state energy $E_0$ 
of the molecule, an effective vertical ionization potential $I_{\rm p}=\vert E_0 \vert + \delta V$ appears in the semiclassical action. With this slight modification, the main oscillatory behavior is now concentrated in the re-defined semiclassical action, and one can safely apply the saddle-point method. There is still one difficulty: the LCAO approximation used to approximate the electronic states in both the molecule and the ion gives rise to ionization matrix elements in Eq.~(\ref{eq:ioniz}) that have a pole close to the saddle-point, since the upper ionzation matrix element from Eq.~(\ref{eq:ioniz}) is proportional to
\begin{eqnarray}\label{eq app: example}
\frac{\mathbf{E}(t)\cdot\mathbf{R}}{2}\sin\left(\frac{[\mathbf{p+A}(t)] \cdot \mathbf{R}}{2}\right)\left(\frac{[\mathbf{p+A}(t)]^2}{2}+\frac{Z^2}{2}\right)^{-2} + \nonumber \\
2\, \mathbf{E}(t)\cdot[\mathbf{p+A}(t)] \cos\left(\frac{[\mathbf{p+A}(t)] \cdot \mathbf{R}}{2}\right) \left(\frac{[\mathbf{p+A}(t)]^2}{2}+\frac{Z^2}{2}\right)^{-3}.
\end{eqnarray}
While for a given time $t$ at which the electronic dipole moment is to be calculated, the equation for the saddle point $t'_{\rm s}$ reads:
\begin{equation}\label{eq app: sad}
 \frac{ [\rm{p}_{\rm s}(t'_{\rm s})+A(t'_{\rm s})]^2}{2} + I_{\rm p} = 0,
\end{equation}
the equation for the pole $t'_{\rm p}$ in the ionization matrix element corresponding to the transition to the gerade state of the ion [see Eq.~(\ref{eq app: example})] is:
\begin{equation}\label{eq app: pol}
 \frac{ [\rm{p}_{\rm s}(t'_{\rm p})+A(t'_{\rm p})]^2}{2} + \frac{Z^2}{2} = 0.
\end{equation}
The parameter $Z$ in Eq.~(\ref{eq app: pol}) is the nuclear charge of the hydrogenic orbitals used in the LCAO approximation. We use $Z=1$. The corresponding value for $Z^2/2=0.5$ in Eq.~(\ref{eq app: pol}) is close to and smaller than the value of $I_{\rm p} = 0.59$ in Eq.~(\ref{eq app: sad}). As a consequence, the two critical points are very close to each other, with the pole having a smaller imaginary part than the saddle point. In this case, the usual saddle-point formula has to be modified accordingly to take into account the presence of the pole. For clarity, let us calculate the contribution of one saddle point $x_{\rm s}$ and one paired pole $x_p$ to the integral:
\begin{equation}
I = \int_0^t dx \, f(x) e^{-i \tilde{S}(x)},
\end{equation}
such that $\tilde S'(x_{\rm s})=0$ and $f(x)$ has a pole of order $3$ in $x_{\rm p}$. (According to Eq.~(\ref{eq app: example}), one has also a second-order pole. Its treatment is entirely similar to that of the $3$rd-order pole.) To this end, we re-write the integral as
\begin{equation}
I = \int_0^t dx \, \tilde{f}(x) \frac{e^{-i \tilde{S}(x)}}{(x-x_{\rm p})^3},
\end{equation}
with $\tilde f(x) = f(x)(x-x_{\rm p})^3$ bounded at the pole. According to the method described in \cite{Bleistein67,Bleistein86}, the first-order asymptotic contribution of the pair of critical points is
\begin{equation}\label{eq app: sp}
I \approx e^{-i \tilde S(x_{\rm s})} \int_{-\infty}^{\infty} dx \left[ 
\tilde{f}_{\rm p} + \frac{\tilde{f}_{\rm s}-\tilde{f}_{\rm p}}{x_{\rm s}-x_{\rm p}} (x-x_{\rm p}) \right] \frac{\exp \left[-\frac{\tilde{S}''_{\rm s}}{2} (x-x_{\rm s})^2\right] } {(x-x_{\rm p})^3}.
\end{equation}
The integrals appearing in Eq.~(\ref{eq app: sp}) are of the form:
\begin{equation}\label{eq app: parab}
\int_{-\infty}^{\infty} dx \frac{\exp(-x^2/2)}{(x-x_0)^k} = 
(-1)^k \sqrt{2\pi} \, e^{-x_0^2/4+ik\pi/2} D_{-k}(ix_0),
\end{equation}
where $x_0 \equiv x_{\rm p}-x_{\rm s}$, ${\rm Im}(x_0)<0$, and $k$ is an integer. Equation (\ref{eq app: parab}) is adapted from the case with ${\rm Im}(x_0)>0$ appearing in \cite{Temme78}. In Eq.~(\ref{eq app: parab}), $D_n(x)$ is the parabolic cylinder function, for which highly accurate numerical routines are available \cite{Zhang96}. Thus, to calculate the dipole momentum from Eq.~(\ref{eq: dip mom2}) at a given time $t$, one needs to find all saddle points and poles with the real part smaller than $t$ and positive imaginary part, and calculate their contribution according to the above. In addition, the contribution from the pole has to be added, according to the Cauchy's theorem. The residue in the pole $x_{\rm p}$ is most simply calculated using directly the simplified expression in Eq.~(\ref{eq app: sp}), which holds in the vicinity of the saddle point (where the pole is situated).
\begin{figure}[!t]
\centering 
\includegraphics[scale=0.5,angle=-90]{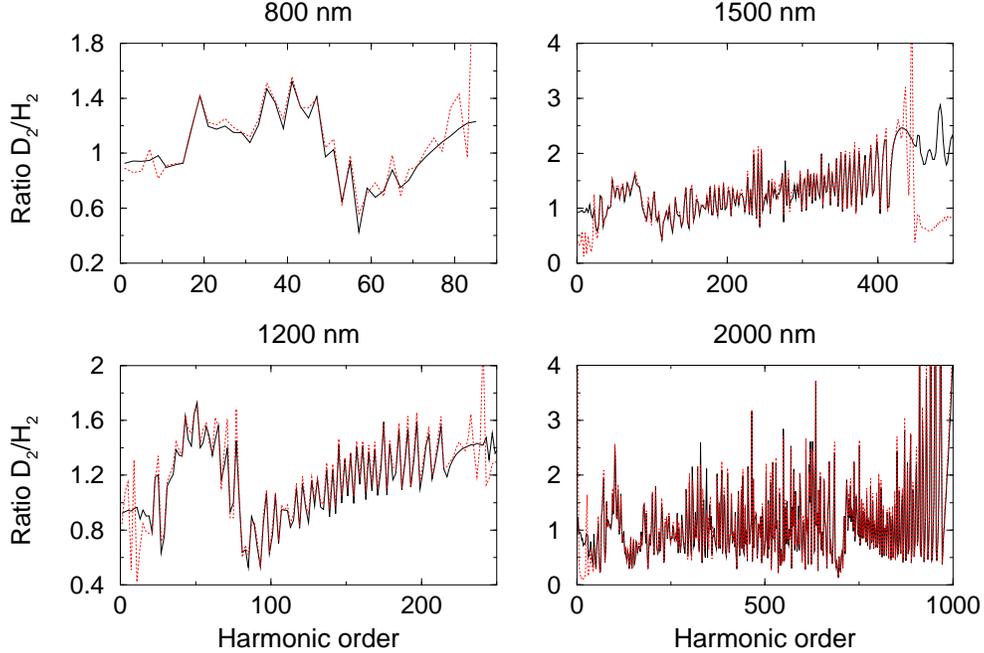} 
\caption{Comparison between the full calculation (black curves) based on Eq.~(\ref{eq: dip mom2}) and by using the saddle-point approximation (red curves). The calculations are for the 1L case.}
\label{figure4}
\end{figure}

The accuracy of the approximation can be seen in Fig.~\ref{figure4}, for various wavelengths. The saddle-point method gives more accurate results, the higher the laser wavelength, which can be seen from the figure. The comparison between exact and saddle-point calculation was carried out for the 1L case using the eigenvalue decomposition \cite{Chirila06} for numerical propagation of the vibrational wave packets, since in this case the calculation is much faster. The eigenvalue decomposition, which can be used only for the 1L case, has been applied only for the purpose of this comparison, while in the rest of this work we use numerical grids. For one laser wavelength, we also checked that the agreement with the saddle-point approximation holds as well for the 2L case ($800$ nm, not shown here). The harmonic-energy region where the agreement should be sought does not include the low-frequency harmonics. The reason is that all the saddle points close to the end point $t$ of the time integration-interval are ignored in our approach. This is because their contribution to the integral can no longer be described by the above procedure. On the other hand, the critical points that are close to the recombination time $t$ will contribute only to the low-frequency part of the harmonic spectrum, which is not the focus of this work. In practice, we ignore the saddle points for which ${\rm Re}(t-t'_s) < T_0/10$
, with $T_0$ the period of one optical cycle. With this in mind, the saddle point method gives very accurate results for the high-energy part of the harmonic spectrum, while reducing significantly the computational time ({\it e.g.}, for a laser wavelength of $1500$ nm, by a factor of $20$).

\end{document}